\documentclass[aps,prl,preprint,superscriptaddress]{revtex4}
\usepackage{graphicx}
\usepackage{bm}
\usepackage{graphicx}
\usepackage{bm}
\begin{document}

\title{Theory and simulation of macromolecular crowding effects on
protein folding stability and kinetics}

\author{Jeetain Mittal}
\email[]{jeetain@helix.nih.gov}
\affiliation{Laboratory of Chemical Physics, National Institute of Diabetes and
Digestive and Kidney Diseases, National Institutes of Health, Bethesda,
Maryland 20892-0520, USA}

\author{Robert B. Best}
\email[]{rbb24@cam.ac.uk}
\affiliation{Department of Chemistry, University of Cambridge, Cambridge, UK}

\date{\today}

\begin{abstract}
We investigate the effect of macromolecular crowding on protein folding, using
purely repulsive crowding particles and a self-organizing polymer model of
protein folding.  We find that the thermodynamics of folding for typical
$\alpha$-, $\beta$- and $\alpha/\beta$-proteins are well described by an
adaptation of the scaled particle theory (SPT). In this approach, the native
state, transition state, and the unfolded protein are treated as effective hard
spheres with radii approximately independent of the size and concentration of
the crowders.  The same model predicts the effect of crowding on the folding
barrier and therefore refolding rates with no adjustable parameters.  
A simple extension of the SPT model, assuming additivity, can also 
describe the behavior of mixtures of crowding particles.
\end{abstract}

\maketitle


In contrast with conventional laboratory experiments conducted under dilute
conditions, protein folding in a cell occurs in a dense environment consisting
of various other macromolecules, commonly referred to as
``crowders''~\cite{Ellis-2003}.  Although the detailed interactions of the
protein with the crowders may be quite complex, the primary physical effect of
macromolecular crowding on the protein folding reaction is to reduce the volume
available to the protein by that occupied by the
crowders~\cite{Minton-1983,Cheung-2005,Cheung-2007a,Minh-2006,Cheung-2007,Zhou-2008r}.
A complete theoretical understanding of the excluded volume effects will
greatly enhance our ability to interpret experiments, as well as all-atom
simulations, and to develop coarse-grained models~\cite{Cheung-2009}, of
concentrated protein solutions.

Several theories have been put forward to predict the effect of crowders,
modeled as impenetrable repulsive particles, on the
folding free energy.  However, these theories provide strongly contrasting
predictions for the quantitative and sometimes even qualitative effects of
crowding.  For example, by treating the folded and unfolded proteins as
effective hard spheres, Minton utilized scaled particle theory (SPT) to
estimate the change in folding free energy as the difference
between the insertion free energy for the folded and the unfolded states. The 
SPT free energy of inserting a hard sphere of 
radius $R$ in a hard-sphere fluid of particle radius $R_\mathrm{c}$
is~\cite{Lebowitz-1964}, 
$\beta F = (3y+3y^2+y^3)\rho + (9y^2/2+3y^3)\rho^2 + 3y^3\rho^3 - {\text
{ln}}(1-\phi_c)$, 
where $\beta=1/k_\mathrm{B}T$, 
$T$ is the temperature, $k_{\text B}$ is the Boltzmann's constant, 
$y=R/R_\mathrm{c}$, $\rho=\phi_c/(1-\phi_c)$, and $\phi_c$ is 
the fluid volume fraction. This
theory predicts a rather strong effect of macromolecular crowding on the
folding free energy~\cite{minton-2005}, with a monotonic increase in stability
with increasing crowder packing fraction $\phi_c$.  

An alternative theory proposed by Zhou also uses SPT for the the effect 
of crowders on the folded protein, but the free energy of the unfolded
protein is calculated using an elegant model of the unfolded chain as a
random walk in the presence of a spherical trap~\cite{zhou-2008}. For a
Gaussian chain of radius of gyration $R_\mathrm{g}$, with $z =
R_\mathrm{g}/R_\mathrm{c}$, the change in free energy is 
$\beta F=3\phi_c z^2+6\pi^{-1/2}\phi_c z - \ln(1-\phi_c)$.
This model predicts a much weaker effect of crowding on the unfolded state free
energy because the unfolded polypeptide can access the void space between the
crowders.  In addition, a maximum in stability is predicted as the crowder
packing fraction is increased. The extension of this theory to binary mixtures
of different size crowders leads to an intriguing conclusion that there will be
an optimum mixing ratio of the two components to achieve maximum protein
stability~\cite{zhou-2008-2}.  One may expect that different crowding theories
will work better under certain conditions of crowder packing fraction and size.
However, the boundaries for the validity of these theories in parameter space
are relatively unknown.

\begin{table}
\caption{\label{parms}Characteristics of the proteins considered.  }
\begin{ruledtabular}
\begin{tabular}{lrrrr}
Protein& $a_\mathrm{N}$ [\AA]  & $a_\mathrm{U}$ [\AA] & $a_{\ddag}$ [\AA] & $R_\mathrm{g}^\mathrm{U}$ [\AA] \\
\hline
prb             & 9.5  &      12.75    &       11.4    & 15.9  \\
proteinG        & 10.6  &      16.0     &       -   &   20.0  \\
TNfn3           & 13.0   &     21.5     &     -           & 28.6  \\
\end{tabular}
\end{ruledtabular}
\end{table}

In this letter, we investigate the appropriate theoretical description of purely
repulsive crowders, using molecular simulations of a coarse grained 
folding model in a bath of crowders. We consider
two-state proteins from the three main protein structural classes: all-$\alpha$ 
(prb~\cite{wang-2004}), all-$\beta$ (TNfn3~\cite{hamill-2000}) and $\alpha/\beta$ 
(protein G~\cite{mccallister-2000}), which are described using a
self-organising polymer model (or ``G\={o}-like'' model) 
~\cite{nymeyer-1998,mickler-2007}. In our model,
each amino acid residue is represented by a single 
particle and a standard procedure is used to build the potential from the 
experimental native-state structure~\cite{karanicolas-2002}. Interactions
between the contacts present in the native state are treated as attractive and
all others as repulsive, an approximation motivated by the funneled nature of
the folding free energy landscape ~\cite{wolynes-2005}. Previous studies have
shown 
that this type of simplified model can indeed capture relevant features 
of protein folding~\cite{shea-2000}, such as mechanism~\cite{levy-2004-2} 
and kinetics~\cite{chavez-2004,best-2005-2}. The repulsive interactions between
a crowder and the protein or other crowders are given by the pair potential
$V(r)=\epsilon[\sigma_{\text {ref}}/(r - \sigma + \sigma_{\text
{ref}})]^{12}$, where $r$ is the distance between particle centers,
$\epsilon=1$ kcal/mol sets the energy scale, $\sigma$ is the hard core overlap
distance, and $\sigma_{\text {ref}}=6$ \AA~ is a reference diameter: for
$\sigma=\sigma_{\text {ref}}$, $V(r)$ reduces to a more familiar form. We
define $\sigma$ between the pair $(i,j)$ as, $\sigma=R_i + R_j$, where $R_i$,
$R_j$ are the radii of either crowder particles $R_\mathrm{c}$, or of the various
protein residues $R_p$.  We use Langevin dynamics simulations 
with a time step of 10 fs and a friction coefficient of 0.2 ps$^{-1}$,
using the BBK integrator~\cite{brooks-1984} in the CHARMM simulation
package~\cite{brooks-1983}. Cubic periodic boundary conditions with a primary
cell size of 100 \AA~ were employed. To speed up equilibration at a given
temperature, we use replica exchange moves every 30 ps between 12 replicas
which are each biased using an umbrella potential of the form $V_i(Q) =
0.5\kappa(Q-Q_i)^2$, where $Q$ is the fraction of native contacts, $0\le Q_i\le
1$ is the target $Q$ value for replica $i$ and $\kappa=300$ kcal/mol is the
force constant. The required thermodynamic information is extracted from
simulations with different umbrella potentials and temperatures using the
weighted histogram method (WHAM)~\cite{ferrenberg-1989,kumar-1992}.

\begin{figure}
\scalebox{0.80}{\includegraphics{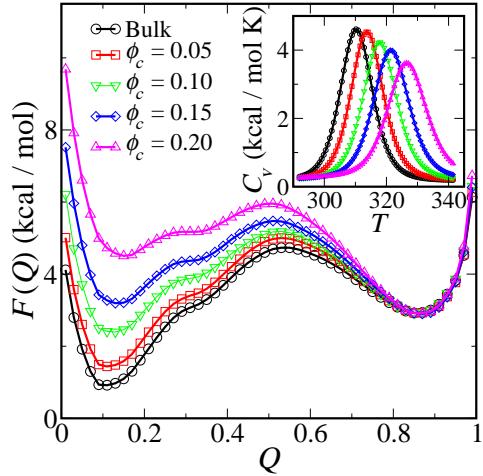}}
\caption{\label{FqCv}{Effect of crowding on folding free energy surface of
protein G.  The potential of mean force along the reaction coordinate $Q$, the
fraction of native contacts, is shown for bulk conditions and with different
crowder packing fractions $\phi_c$ for crowder size $R_\mathrm{c}=8$ \AA at 
320 K. The
curves have been shifted by an arbitrary constant to match the free energy of
the folded state. 
Inset: Heat capacity $C_\mathrm{v}(T)$. 
}}
\end{figure}

For the model proteins considered here, $Q$ is nearly an 
optimal coordinate for identifying transition states and also 
capturing the dynamics of protein folding with a diffusive Markovian model
~\cite{best-2005-2,best-2006-2,mittal-2008}. 
Here, we project folding in our systems onto $Q$, and use it 
to identify transition states as well as unfolded and folded 
free energy minima. To estimate the crowding-induced changes on the 
folding free energy surface, we construct the potential of mean force (PMF) 
as a function of $Q$, defined as $\beta F(Q)=-\ln[P(Q)/\Delta Q]$, 
where $P(Q)$ is the equilibrium probability of observing configurations between
$Q$ and $Q+\Delta Q$.
Figure~\ref{FqCv} shows representative free energy profiles in bulk
($\phi_c=0$) and under crowded conditions ($\phi_c > 0$, $R_\mathrm{c}=8$) for
protein G. The destabilization of the unfolded state due to the presence of
crowders is clearly visible from an upward shift in the curves near the
unfolded basin ($Q\approx 0.1$). For high crowding packing fractions ($\phi_c >
0.10$), there is also a slight destabilization of the transition state ($Q
\approx 0.5$) with respect to the folded basin ($Q \approx 0.9$).
Fig.~\ref{FqCv} (inset) shows the heat capacity $C_\mathrm{v}$ curves for the
same state points, as obtained from the WHAM analysis. The maximum in
$C_\mathrm{v}$ is defined as the folding temperature $T_\mathrm{f}$. The shift
in $T_\mathrm{f}$ toward higher temperatures with increasing $\phi_c$
demonstrates the stabilizing effect of the crowders on the folded protein
relative to unfolded.

\begin{figure}
\scalebox{0.70}{\includegraphics{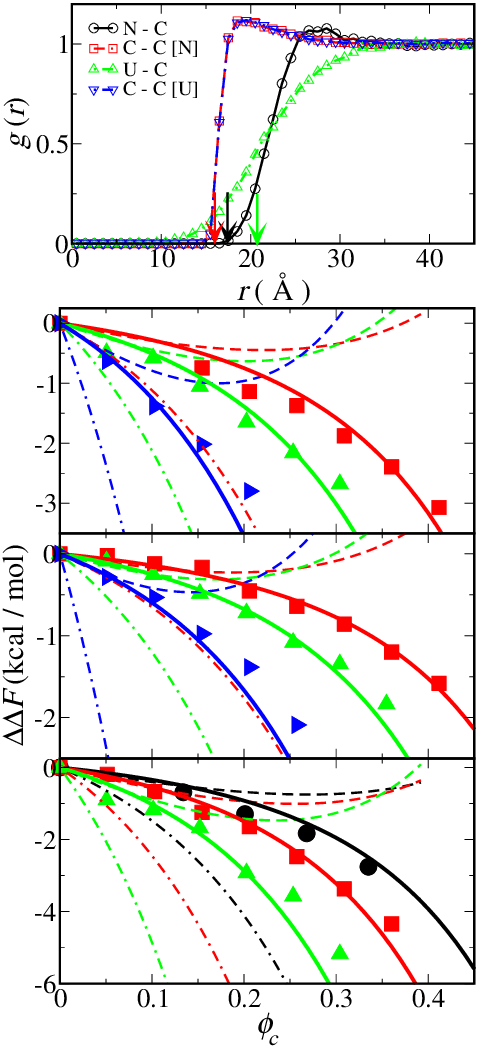}}
\caption{\label{ddF}{Shift in protein stability with crowder size and packing
fraction. Top panel: pair distribution functions $g(r)$ between crowders (C)
and the native (N) and unfolded (U) protein (C-C [U] is the C-C $g(r)$ in the
presence of U), for prb with $R_c = 8$ \AA, $\phi_c=0.05$. Arrows indicate
corresponding sums of hard sphere radii. 
Lower panels:
change in folding free energy with crowding: \textit{upper}: protein G,
\textit{middle}: prb, \textit{lower}: TNfn3. Symbols: simulations
(errors smaller than symbol size) for $R_\mathrm{c}=20$ \AA~ (circle), 16 \AA~ (square), 
12 \AA~ (triangle up), and 8 \AA~ (triangle right); dashed lines: predictions
from Zhou theory~\cite{zhou-2008}; dot-dash lines: predictions from
Minton theory with $a_\mathrm{U}=\sqrt{5/3}R_\mathrm{g}^\mathrm{U}$; 
solid line: result of using a smaller hard sphere for the unfolded
state, motivated by the form of the U-C $g(r)$.}}
\end{figure}

Figure~\ref{ddF} shows the effect of crowding on the free energy of folding,
$\Delta\Delta F=\Delta F_\mathrm{U-N}(\phi_c=0) - 
\Delta F_\mathrm{U-N}(\phi_c)$, at $T=320$ K for prb and protein G and 
$T=300$ K for TNfn3, 
where $\Delta F_\mathrm{U-N} = -k_{\text B}T {\text {ln}} 
(\int_0^{Q_{\ddag}} {\text e}^{-\beta F(Q)}dQ/
\int_{Q_{\ddag}}^1 {\text e}^{-\beta F(Q)}dQ)$, 
is the difference between the unfolded $F_{\text {U}}$ and 
native $F_{\text {N}}$ free energy, $Q_{\ddag}$ is the location of the transition 
state along $Q$.
We find a
monotonic increase in stability with the packing fraction $\phi_c$ for all
proteins and crowder sizes. For a given $\phi_c$, the smaller crowders are more
strongly stabilizing because there are
effectively fewer voids of the size of the protein.
The prediction of Zhou's theory using the folded protein radius $a_\mathrm{N}$
and unfolded protein radius of gyration $R_\mathrm{g}^\mathrm{U}$ computed from
the simulations (Table~\ref{parms})
gives excellent agreement with the simulation data in the limit $\phi_c \to 0$
(Figure~\ref{ddF}). For smaller crowders and higher packing fractions, however,
the theory predicts too small a stabilization, most likely due to the neglect of
excluded volume in the unfolded chain.

On the other hand, the Minton theory predicts a stabilization much greater than
found in the simulations over the full range of $\phi_c$. This theory treats
the unfolded protein (with radius of gyration $R_\mathrm{g}^\mathrm{U}$) as an 
``equivalent hard sphere'' of uniform density,
i.e. with radius $\sqrt{5/3}R_\mathrm{g}^\mathrm{U}$.  However, as indicated by the pair
distribution function $g(r)$ between the unfolded protein and the crowders
(Figure~\ref{ddF}), the unfolded protein is quite soft, so that a smaller
choice of hard sphere radius may be more appropriate.  Remarkably, we are able
to fit the data over the full range of $\phi_c$ and crowder size (which varies
by a factor of 2) to the SPT
model by using the radius of the unfolded state $a_\mathrm{U}$ as a single
adjustable parameter.  This radius is indicated on the plot of $g(r)$ in
Figure~\ref{ddF}. At high $\phi_c$ the smallest crowders 
are better able to penetrate
the protein and the fit could be marginally improved by using a slightly smaller
$a_\mathrm{U}$. This treatment of the unfolded protein as a hard sphere is
motivated by the highly successful Asakura-Oosawa theory~\cite{asakura-1954} of
polymer-colloid mixtures so that soft sphere interactions can be used between
the polymer and crowder~\cite{louis-2000} which in turn can be mapped onto a
hard-sphere system. In our case, we find that the effective hard-sphere radius
$a_\mathrm{U}$ of the unfolded chain is
$\approx$ 80$\%$ of $R_\mathrm{g}^\mathrm{U}$.

\begin{figure}
\scalebox{0.70}{\includegraphics{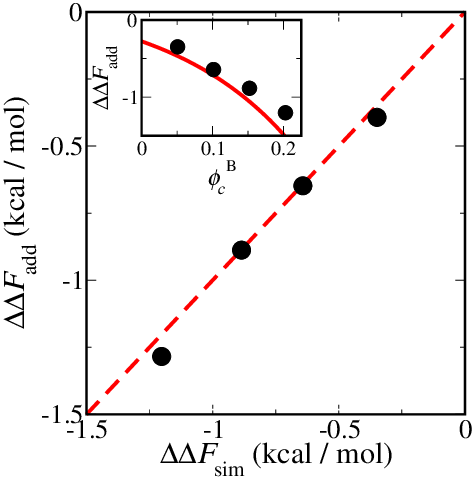}}
\caption{\label{pred}{Scaled particle theory prediction for binary 
mixtures of 8 \AA~ and 12 \AA~ crowders.  With the packing fraction of particle
type A fixed to $\phi_c^A=0.05$ and the fraction of particle type B, 
$\phi_c^B=0.05$ - 0.20, the change in folding free energy of prb from 
simulations $\Delta\Delta F_{\text {sim}}$ is shown as a function of 
change in folding free energy 
$\Delta\Delta F_{\text {add}}$ 
from the additivity {\em ansatz} (circle) as discussed in the text.
The inset shows simulation data (circle) along with SPT model predictions (line) using the additive model. 
}}
\end{figure}

We have also studied mixed macromolecular crowding with binary A:B and ternary
A:B:C mixtures for the $\alpha$-helical protein prb. For A:B mixtures, keeping
the packing fraction of the type ``A'' crowders (radius 8 \AA) fixed at
$\phi_c^A=0.05$, we varied the fraction of type ``B''
crowders (radius 12 \AA). The simulation results for the change in
folding free energy for different mixtures are shown in Figure~\ref{pred}.
To test whether the effect of mixed crowding is simply additive and can be 
easily estimated from the pure crowder simulations, we also calculate the 
change in folding free energy from the following additivity {\em ansatz}, 
$\Delta\Delta F_{\text {add}}(\phi_c^1,\phi_c^2,...\phi_c^N)=
\sum_i x_i \Delta\Delta F_i(\sum_i \phi_c^i)$, where index $i$ runs over 
$N$ different types of crowding particles, 
$x_i=\phi_c^i/\sum_j\phi_c^j$ is the fraction 
of crowder type $i$ in the mixture, and $\phi_c^i$ is the volume 
fraction of $i$. 
The predictions of this additive model $\Delta\Delta F_{\text {add}}$ for A:B
mixtures are in extremely good agreement with the simulation results 
$\Delta\Delta F_{\text {sim}}$ as shown in 
Fig.~\ref{pred}. Moreover, for an A:B:C mixture of crowder radii 8, 12, 
16 \AA~ with volume fraction $\phi_c^i= 0.05$ of each component, the 
agreement between the simulation result (0.55 kcal/mol) and the 
additive model (0.54 kcal/mol) is remarkably good. Therefore, it should 
be possible to estimate the effect of mixed crowding due to several 
types of repulsive crowder by utilizing single crowder results. 
Applying additivity to our SPT model predictions for single crowders
also provides reasonable estimates for $\Delta\Delta F$ (Figure \ref{pred}
inset).
Our results and the additive model for mixed crowding do not predict
an optimal mixing ratio as expected from a previous
theory~\cite{zhou-2008-2}.  Indeed, for a mixture of crowders at given
$\phi_c$, the greatest stabilization will occur when all the crowders are of
the more stabilizing (smaller) type. 

\begin{figure}
\scalebox{0.75}{\includegraphics{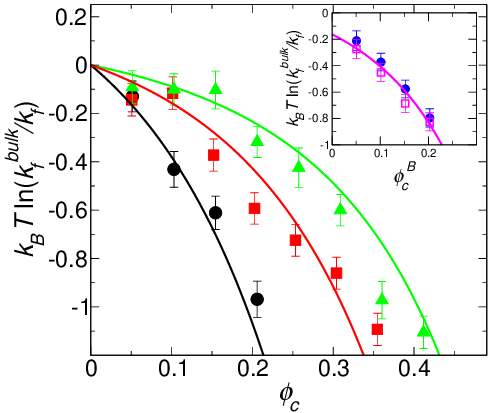}}
\caption{\label{kin}{Influence of crowding on folding kinetics. 
The effect of crowding on the folding barrier height of prb is estimated using
$k_\mathrm{B}T\ln(k_\mathrm{f}^\mathrm{bulk}/k_\mathrm{f})$ for crowders of
radius 8 \AA~ (circle), 12 \AA~ (square) and 16 \AA~
(triangle up); lines are the SPT predictions. \textit{Inset}:
Folding rates with binary crowder mixtures (A: 8 \AA~ radius, B: 12 \AA~
radius, $\phi_c^A = 0.05$). Results from direct simulations (circle)
are compared with the predictions based on simulations with
a single crowder size (square) and from SPT (line).
}}
\end{figure}

From our explicit dynamical model of folding we are able to estimate
folding rates directly from mean first passage time calculations. Starting from
at least 400 different initial coordinates drawn from an equilibrium unfolded ensemble
($Q=0.2$) at a given packing fraction, we calculate the average time
$\tau_\mathrm{f} = 1/k_\mathrm{f}$ taken to reach the folded state ($Q>0.9$).
We estimate the change in barrier
height ($\Delta F_{\ddag\mathrm{-U}} = F_{\ddag} - F_{\text {U}}$;
$F_{\ddag}=F(Q_\ddag)$) 
with crowding as 
$k_\mathrm{B}T\ln(k_\mathrm{f}^\mathrm{bulk}/k_\mathrm{f})$, which is
justified if the position of the folding barrier and the diffusion
coefficient along the reaction coordinate are unchanged (Figure~\ref{kin}). 
The folding barrier height decreases monotonically with increasing $\phi_c$:
indeed, it is possible to predict the change in rates using SPT with no 
further adjustable 
parameters: we calculate the transition state radius $a_{\ddag}$
directly from the $Q$ umbrella simulations ($Q=0.55$), and
estimate the change in barrier height from SPT, using $a_{\ddag}$ in place of
$a_\mathrm{N}$. The effect of mixed macromolecular crowding on kinetics
can also be obtained from single crowder simulations by assuming
additivity (Figure~\ref{kin}, inset).

In summary, we find that scaled particle theory provides an accurate
description of the effect of macromolecular crowding on both folding stability 
and rates.
The effect of purely repulsive crowders can be well-approximated
over a wide range of crowder sizes and packing fractions by treating the
unfolded state as a hard sphere of fixed radius $a_\mathrm{U}$. In all the
cases considered,
we find $a_\mathrm{U}\approx0.8R_\mathrm{g}^\mathrm{U}$. Our results have
several important consequences. For crowders of a single size, folding rate and
stability will increase with increasing packing fraction monotonically under
the relevant physiological conditions. When considering mixtures of purely
repulsive crowders of different sizes the effects of crowding are additive, and
the most stabilizing composition will consist completely of the smallest
crowder.  Therefore any maximum in stabilization as a function of the ratio of
various components at a fixed total packing fraction implies the existence of
attractive interactions with at least one of the crowders~\cite{du-2006}.

\begin{acknowledgments}
We are grateful to Dr. Attila Szabo for several helpful discussions.  This
research was supported by the Intramural Research Program of the NIH, NIDDK.
J.M.  thanks Dr. Artur Adib for supporting a postdoctoral fellowship. R.B. is
supported by a Royal Society University Research Fellowship.  We gratefully
acknowledge the Human Frontier Science Program short-term fellowship to J.M.
for which this work was originally proposed.  This study utilized the
high-performance computational capabilities of the Biowulf PC / Linux cluster
at the National Institutes of Health, Bethesda, MD (http://biowulf.nih.gov),
Theory Sector computing resources at Cambridge, and the NSF TeraGrid resources
provided by TACC.
\end{acknowledgments}


\end{document}